# Machine Learning Applications in Traumatic Brain Injury: A Spotlight on Mild TBI

Hanem Ellethy,[1] Shekhar S. Chandra[1], and Viktor Vegh[2,3]

[1] School of Electrical Engineering and Computer Science, University of Queensland, QLD, Australia.
[2] The Centre for Advanced Imaging, Australian Institute for Bioengineering and Nanotechnology, University of Queensland, QLD, Australia.
[3] ARC Training Centre for Innovation in Biomedical Imaging Technology, QLD, Australia.

`h.elwaseif@uq.edu.au`

## Abstract

Traumatic Brain Injury (TBI) poses a significant global public health challenge, contributing to high morbidity and mortality rates and placing a substantial economic burden on healthcare systems worldwide. The diagnosis of TBI relies on clinical information along with Computed Tomography (CT) scans. Addressing the multifaceted challenges posed by TBI has seen the development of innovative, data driven approaches, for this complex condition. Particularly noteworthy is the prevalence of mild TBI (mTBI), which constitutes the majority of TBI cases where conventional methods often fall short. As such, we review the state-of-the-art Machine Learning (ML) techniques applied to clinical information and CT scans in TBI, with a particular focus on mTBI. We categorize ML applications based on their data sources, and there is a spectrum of ML techniques used to date. Most of these techniques have primarily focused on diagnosis, with relatively few attempts at predicting the prognosis. This review may serve as a source of inspiration for future research studies aimed at improving the diagnosis of TBI using data driven approaches and standard diagnostic data .

## 1 Traumatic Brain Injury

Traumatic Brain Injury (TBI), often referred to as the silent epidemic [1], is a disruption of normal brain function due to external forces [2]. Common causes include motor vehicle accidents, falls, blunt force trauma, assaults, self-inflicted injuries, and other incidents [3]. The Glasgow Coma Scale (GCS) scores TBI severity in categories of mild (13–15), moderate (9–12), or severe (<9) [4]. TBI is a significant global health challenge, contributing to high rates of morbidity and mortality and placing a substantial economic burden on healthcare systems [5]. Reports estimated 55.5 million active TBI cases worldwide in 2016 [6], 2.78 million in the U.S. in 2014 [7], and 2.5 million in the European Union [8]. Mild TBI (mTBI) accounts for over

70% of all TBI cases in the U.S. [9], [10], and around 91% globally [11]. Despite associated with a 'mild' classification, mTBI can lead to significant cognitive impairments, emotional and behavioural changes, and an increased risk of long-term neurodegenerative diseases [12]–[16], all of which contributing to a major public health concern [17].

In the Emergency Department (ED), the primary place for acute TBI evaluation, physicians assess patients using clinical and neurological examinations (often GCS) to determine injury severity. Computed Tomography (CT) scans are the standard neuroimaging tool for any suspected TBI to rapidly identify hemorrhages, hematomas, and skull fractures [18]. However, clinical assessments and CT scans can lack the sensitivity needed for a definitive TBI diagnosis [9]. In such cases, Magnetic Resonance Imaging (MRI) may be used for follow-up or when patients do not show expected improvements post-injury to obtain a better assessment of soft tissue damage [19].

TBI evaluations impose a significant burden on EDs. The Centres for Disease Control and Prevention (CDC) in the U.S. reported approximately 2.87 million TBI-related ED visits, hospitalizations, and deaths in 2014, representing a 53% increase from 2006 and accounting for about 3.6% of annual ED visits [3]. In 2019, approximately 15% of U.S. high school students reported sports-related concussions (or mTBI) in the preceding year, and in 2021, there were over 69,000 TBI-related deaths [20]. Furthermore, nearly 3.9 million CT scans are performed each year to assess TBI patients, with around 91% of them labelled negative, and only 22% of the TBI-diagnosed patients have positive CTs [9]. The high volume of negative head CT scans conducted in EDs not only requires substantial resources and time [9], but also increases exposure to ionising radiation (i.e., X-rays) [21]. This has piqued interest in using data driven approaches to predict the need for a CT scan in individuals assessed in EDs.

TBI is a highly variable and patient-specific condition due to the spectrum of possible injuries to the head of a person. Therefore, collecting extensive TBI-related data in the early phases after injury is crucial to better understand and manage TBI. Several clinical trials, including the International Mission for Prognosis and Analysis of Clinical Trials (IMPACT), Transforming Research and Clinical Knowledge in Traumatic Brain Injury (TRACK-TBI), and CENTER-TBI, were initiated between 2009 and 2011. These trials, in addition to routine clinical information, incorporated blood biomarkers and imaging techniques to improve patient prognosis. TRACK-TBI was established to develop, evaluate, and refine Common Data

Elements for TBI, and it serves as a substantial database of information for researchers and investigators [22]. These clinical trials aim to gather a wealth of data while managing and extracting relevant information from this data pose significant challenges. Clinical decision rules play a crucial role in TBI management and are derived using data from these clinical trials. They help identify TBI based on clinical attributes and guide the use of head CT scans. The Paediatric Emergency Care Applied Research Network (PECARN) is one such rule, identifying low-risk TBI in children, allowing for safe discharge without CT scans [23]. The Canadian Assessment of Tomography for Childhood Head injury (CATCH) rule is used to predict significant head injuries in children and then determine the need for a CT scan [24]. For adults, the Canadian CT Head Rule (CCHR) identifies mTBI patients requiring CT scans and assesses neurosurgical risk [25]. While these rules exhibit high sensitivity, they lack specificity [26], which in turn result in only minimal reductions in CT scans [9].

The field of TBI research faces considerable challenges due to the heterogeneity of the condition in terms of severity, causes, treatments, pathology, origins, and outcomes. Currently, there exist substantial gaps between the available information and our understanding of TBI diagnosis and treatment. The current shortcomings encompass the accurate diagnosis of mTBI, the classification system employed, effective treatments, and predictions of disease progression. It is thus prudent to enhance the accuracy, efficiency, timeliness, and cost-effectiveness of TBI diagnosis and prognosis to reduce the burden on EDs tasked with TBI evaluations. Data driven methods that harness information from patient history, clinical examinations, neuroimaging results, medication records, hospital admissions, and more, are needed to advance the diagnosis and prognosis of TBI. Automated techniques, namely Machine Learning (ML) and Deep Learning (DL), offer new opportunities in improving TBI diagnosis and prognosis.

## 2   Machine Learning

Artificial Intelligence (AI) as enabled by computer science focuses on creating systems capable of tasks that typically require human intelligence. These tasks include problem-solving, pattern recognition, language understanding, and decision-making. AI emulates human cognitive functions with expanding applications in the medical sector [27].

ML within the broader area of AI differs from traditional programming by enabling systems to learn and improve from experience. ML frameworks in the form of algorithms executed on computers learn from data, infer patterns, and make decisions with minimal human input [28]. Deep Learning (DL) constructs, considered as ML working with large data of high complexity, were inspired by the neuronal architecture of the human brain leading to the concept of Artificial Neural Networks (ANN) [29]. The simplified encapsulation of DL within ML and AI is depicted in Figure 1.

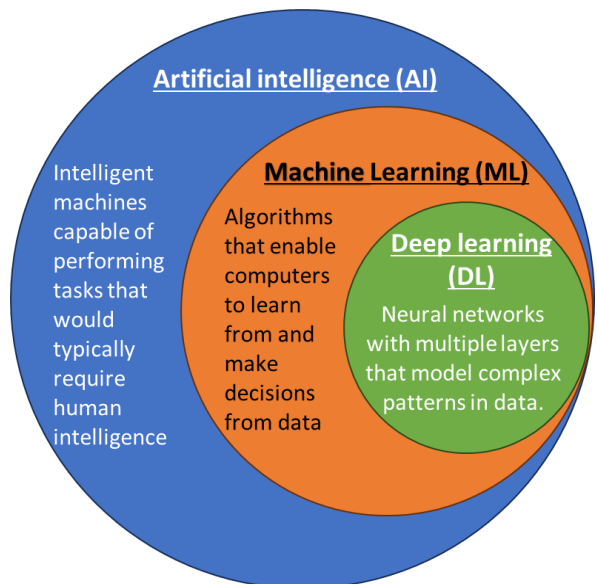

**Figure 1: The visual representation of the relationship between AI, ML, and DL.**

ML continues to gain traction in the medical field due to its ability to work with intricate datasets [30], and it stands apart from traditional statistical models and conventional healthcare practices for several reasons. Firstly, it can automatically identify patterns in data and improve its performance through experience when tasked with specific functions like classification or prediction, thus eliminating the need to explicitly program each task. Secondly, ML can extract meaningful features from complex and diverse data using a range of advanced models, such as those involved in DL, enabling it to make meaningful inferences from complex information. Thirdly, ML is adept at handling vast volumes of data, even when patient records are in the order of hundreds of billions, and still identify the key features in the data [28], [29], [31], [32]. ML algorithms are often broadly categorized into supervised and unsupervised approaches. Supervised algorithms learn from labelled data, establishing a relationship between inputs (features) and outputs (labels). Once trained, these algorithms are used to predict or classify new input data. Unsupervised algorithms are tasked to uncover hidden patterns within unlabelled data, the outcome of which is the prediction of key features in the data [33]. Together, these advantages provide ML approaches opportunities within TBI, where ultimately, patient specific predications are needed.

The shallow-ANN, considered as a traditional ML technique, has a distinct difference compared to recent DL implementations. This neural network consists of only one hidden layer between the input and output layers. An important aspect here is feature engineering, which then defines the input layer, and the choice of feature can significantly impact model performance [34]. The Random Forest (RF), considered as a traditional ML method as well, performs ensemble learning through a multitude of decision trees necessary to achieve a robust model. Each tree within an RF makes its own class prediction, and the class with the most votes is considered as the correct prediction. An RF can be used as a classification tool or for feature selection [35]. By using RF for feature selection, one can enhance model performance, reduce complexity, and increase interpretability, making it a versatile tool for ML applications [36]. For these methods feature engineering involves selecting and transforming raw data into features that accentuate the underlying patterns in the data. The process often relies on specific domain knowledge and is particularly important when working with smaller datasets where instances are lacking for statistical power. Normally, it is straightforward to interpret the meaning of the feature since it was created to highlight a specific aspect of the raw data. This trackability is important in healthcare, where decisions need to be explainable and properties of the data are used to inform clinical decision making [37].

DL models have been shown to excel at learning complex patterns directly from raw data without a need for manual feature engineering [38]. This attribute makes DL highly effective for handling large and intricate datasets. Within the DL context, the Convolutional Neural Network (CNN) is adept at processing data with a grid-like topology, such as images, making them highly effective for tasks involving image recognition, image classification, and object detection in images [39]. A CNN typically consists of multiple layers that iteratively and adaptively learn spatial hierarchies of features from input images. A CNN consists of convolutional, pooling, and fully connected layers. The convolutional layers apply a convolution operation to the input, passing the result to the next layer. This process involves the use of filters or kernels to extract features such as edges, textures, and more complex patterns in deeper layers. Pooling layers, usually placed after a convolutional layer, reduce the spatial dimensions (width and height) of the input volume before processed by the next convolutional layer, essentially leading to a stepwise dimensionality reduction of the input image. The fully connected layers, usually placed at the end of the network, perform classification based on the features extracted by the convolutional and pooling layers. The strength of a CNN lies in its ability to learn feature representations automatically, without the need for manual feature extraction. This characteristic makes a CNN particularly suited for tasks where the feature set may be too complex to be hand-crafted. Over the years, CNNs have set benchmarks in a wide range of image processing tasks and continue to be at the forefront of the field of computer vision. This foundation has enabled CNNs to excel in various application areas by learning hierarchical feature representations from imagery data. This ability to discern and utilize intricate visual features makes them exceptional at tasks involving visual analysis [38].

The difference between the traditional ML and DL approaches is depicted in Figure 2. The distinctions highlighted are fundamental in selecting the most suitable approach based on the size and complexity of the dataset at hand, as well as the importance of interpreting the decisions made by an ML model. Interpretation of raw data features is particularly important in specialized fields like the diagnosis of TBI, where the data feature used to make a decision can significantly impact the effectiveness and reliability of the diagnosis. Table 1 summarizes the most used ML models include the model's name, whether it's a traditional ML or DL model, if it's primarily used for feature extraction, and a concise description of each.

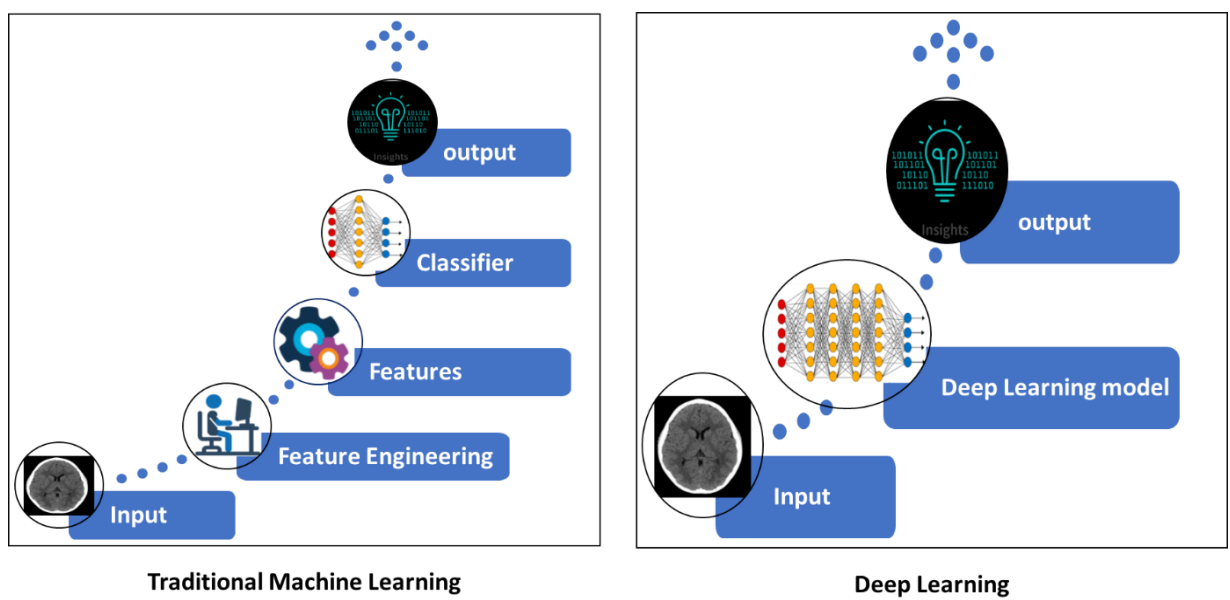

**Figure 2: The difference between the traditional ML and DL approaches.**

**Table 1**: Summary of the most used machine learning and deep learning models.

| Model | Type | Classifier | Feature Extraction | Description |
|---|---|---|---|---|
| ANN | ML | Yes | No | A basic neural network with fewer hidden layers, used for simple pattern recognition tasks [40]. |
| LR | ML | Yes | No | A statistical model used for binary classification problems [41]. |
| E-NB | ML | Yes | No | Combines predictions from multiple Naive Bayes classifiers to improve accuracy [42]. |
| DT | ML | Yes | No | A tree-like model used for classification and regression tasks [43]. |
| KNN | ML | Yes | No | A simple, non-parametric algorithm used for classification and regression [44]. |
| SVM | ML | Yes | No | A powerful classifier that works well on both linear and non-linear problems [45]. |
| XGB | ML | Yes | No, but it assesses feature importance | An efficient implementation of gradient boosting for classification and regression [46]. |
| RF | ML | Yes | No, but it assesses feature importance | An ensemble learning method for classification and regression that constructs multiple decision trees [47]. |
| WKHC | ML | No | Yes | An advanced clustering method using weighted K-means and histogram techniques [48]. |
| LDA | ML | No | Yes | A method used in statistics, pattern recognition, and machine learning for finding linear combinations of features [49]. |
| SIFT | ML | No | Yes | An algorithm in computer vision to detect and describe local features in images [50]. |
| NCA | ML | No | Yes | A machine learning algorithm for distance metric learning [51]. |
| CNN | DL | | Yes | Primarily used for image processing, utilizes convolutional layers to extract features [52]. |
| U-Net | DL | | Yes | A distinctive U-shaped convolutional network architecture for fast and precise segmentation of images [53]. |
| RADnet | DL | | Yes | A network that combines recurrent neural networks with attention mechanisms and dense connectivity [54]. |
| AE | DL | | Yes | A type of deep neural network used to learn efficient coding of unlabelled data [55]. |
| VGG | DL | | Yes | A deep neural network known for its depth and architecture used in image recognition tasks [56]. |
| ResNet | DL | | Yes | A deep neural network with skip connections, allowing it to learn residual functions [57]. |

*ML: Machine Learning, DL: Deep Learning, ANN: Artificial Neural Network, LR: Logistic Regression, E-NB: Ensemble of Naive Bayes Classifiers, DT: Decision Tree, KNN: K-Nearest Neighbour, SVM: Support Vector Machine, XGB: Extreme Gradient Boosting, RF: Random Forest, WKHC: Weighted K-means Histogram-based Clustering, LDA: Linear Discriminant Analysis, SIFT: Scale Invariant Feature Transformation, NCA: Neighbourhood Component Analysis, CNN: Convolutional Neural Network, RADnet: Recurrent Attention Dense Network, AE: Autoencoder, VGG: Visual Geometry Group, ResNet: Residual Neural Network.*

# 3 Applications of Machine Learning in TBI

We have categorized this review by the standard diagnostic data, encompassing both clinical information and CT scans, for a range of ML tasks related to TBI. These tasks include diagnosis and prognosis of TBI, as well as segmentation and classification of CT images. For mTBI, we additionally consider studies that utilized other modality data like functional connectivity, EEG and MRI, to delve into the specific ML applications employed for mTBI.

## 3.1 Machine Learning using Clinical Information

Table 2 lists the studies that have effectively used clinical data for TBI diagnosis and prognosis. Provided are the variety of ML models employed, whether the CT scan report was added to the clinical data, the spectrum of injury severities covered, and the specific goals of each study (i.e., diagnostic, or prognostic purpose). A notable point to make is that these studies predominantly used traditional ML models. From the 20 studies listed, 19 applied traditional ML methods and one a DL method (i.e., deep-ANN). Studies are equally spread between diagnosis and prognosis of TBI. The larger focus towards enhancing the CT decision-making process following mTBI, which influences the diagnosis process and aid choices for tailored treatment strategies. In terms of prognostic models, the aim has been to predict outcome after TBI diagnosis. Here, clinical images are not used directly, instead in five studies the actual CT report was coded as additional clinical information. In general, models incorporating the CT report have outperformed those that did not consider imaging findings in the ML frameworks, suggesting that the more comprehensive the information the better the result (as measured using metrics of accuracy, specificity, and AUC).

In these studies the ML models consider a combination of clinical and demographic data, with or without CT reports, and appear to be constructively informing for TBI cases [58]. An area of interest has been deciding the need for a CT scan in mTBI, especially in pediatric cases. Klement et al. employed a naive Bayes ensemble model for this task and achieved sensitivity of 83% and a specificity of 74% [42], while Miyagawa et al. reported a high accuracy of 95% and an AUC of 85% using a DT model [43]. In a substantial cohort of 14,983 patients from the PECARN study, a deep-ANN was employed to predict the need for CT scans, in comparison to the PECARN clinical rules [59]. This approach achieved a sensitivity of 94% and a specificity of 98%, as opposed to the PECARN rules which had a sensitivity of 100% and specificity of 54%. In adult populations, Molaei et al. utilized an RF model to identify TBI patients in need of CT scans (sensitivity = 82%; specificity = 76%) [60]. In an elderly cohort Dusenberry et al. applied an ANN model to predict acute CT findings (sensitivity = 98%; specificity = 89%) [61]. The integration of CT report data further enhanced the diagnostic capabilities, as evidenced by our research that showcased the potential of both shallow and deep ANN models in diagnosing mTBI (specificity = 100%; sensitivity = 99%) [40]. In terms of acute management across various TBI severities, Hale et al. employed a shallow-ANN to predict clinically relevant TBI outcomes (sensitivity = 100%; specificity of 60%) [62]. Additionally, M. Zhang et al. assessed multiple models for predicting acute functional outcomes at hospital discharge, with the RF model performed the best (sensitivity = 75%; specificity = 81%) [63]. Li et al. focused on the prognosis of brain contusion and laceration post-surgery using LR and ANN [87]. They were particularly interested in predicting cerebral contusion after surgery (AUC = 82%). These advancements highlight the critical role of ML models in enhancing the precision and effectiveness of TBI care, catering to diverse patient groups and injury severities. They also highlight that advances in ML frameworks also lead to improved predictions.

Table 2: Summary of the studies using clinical data with machine learning in TBI.

| Study | Year [Ref] | Model | CT Report | Severity | objective |
|---|---|---|---|---|---|
| Klement et al. | 2012 [56] | E-NB | | Mild | Diagnosis |
| Molaei et al. | 2016 [71] | CSRF | No | | |
| Dusenberry et al. | 2017 [72] | ANN | | | |
| Ellethy et al. | 2021 [2] | ANN | Yes | | |
| Ellethy et al. (1) | 2022 [1] | Deep-ANN | No | | |
| Miyagawa et al. | 2023 [22] | DT | | | |
| Hale et al. | 2019 [73] | ANN | No | Mild to severe | |
| M. Zhang et al. | 2023 [74] | XGB, RF, ANN, … | | | |
| Bittencourt et al. | 2021 [58] | SVM | No | Mild | Prognosis |
| Dabek et al. | 2022 [75] | ANN, SVM, LR, … | | | |
| Fonseca et al. | 2022 [76] | XGB, RF, ANN, … | Yes | Mild to severe | |
| Say et al. | 2022 [77] | XGB & RF | | | |
| Van der Ploeg et al. | 2016 [55] | LR, SVM, RF, ANN | Yes | Moderate to severe | |
| Hale et al. | 2018 [78] | ANN | | | |
| Farzaneh et al. | 2021 [59] | XGB | | | |
| Minoccheri et al. | 2022 [79] | TFNN | | | |
| Yang et al. | 2021 [80] | Ada, RF, ANN, … | No | | |
| Wang et al. | 2022 [81] | LR, XGB | | | |
| Li et al. | 2022 [87] | LR, ANN | | | |
| Z. Zhang et al. | 2023 [82] | LR, XGB, LGBM, … | | | |

*CSRF: Cost-Sensitive Random Forest , TFNN: Tropical geometry-based Fuzzy Neural Network, Ada: adapting boosting, LGBM: Light Gradient Boosting Machines*.

The addition of CT report information may be useful to the clinical input data for the ML algorithm has been found to improve predictions, however not all studies on the prognosis of TBI have incorporated this type of information due to availability. In an elderly cohort, using only clinical TBI assessment data from UPFRONT and ReCONNECT studies, the SVM classification produced an AUC = 80% based on self-reported post-TBI outcomes [45]. Interestingly, after careful assessment of the data, the authors found post-injury neck pain, irritability, and forgetfulness as determinates of incomplete recovery after mTBI diagnosis. In mTBI, Dabek et al. focused on using the extended clinical information to assess multiple ML models for predicting post-injury mental health [64]. Their ANN implementation achieved 88% accuracy. Addressing moderate-to-severe TBI, Yang et al. explored the risk of coagulopathy. They adapted an existing boosting model, leading to prediction accuracy of 92% [69]. Subsequently, Z. Zhang et al. utilized light gradient boosting machines to predict mortality in moderate-to-severe TBI cases (accuracy = 95%) [71]. This work was in agreement with a parallel study by Wang et al., who found Extreme Gradient Boosting (XGB) to predict survival after subarachnoid hemorrhage with an AUC of 95% (versus 77% with LR) and unfavourable outcome AUC of 96% (versus 83% with LR) [70]. In mild-to-severe TBI, Say et al. RF was used to predict functional independence measure scores post-rehabilitation (they showed only a low loss function value) [66]. Fonseca et al., on the other hand, focused on predicting hospital discharge mortality in a similar cohort. They employed XGB and achieved an AUC of 91% [65], suggesting XGB can achieve high precision for prognostic modelling. For predictions of unfavourable outcomes in moderate-to-severe TBI, Minoccheri et al. implemented an ANN enhanced with tropical geometry-based Fuzzy logic (accuracy = 72%) [68], and Farzaneh et al. applied XGB to predict six-month functional outcomes (accuracy = 75%) [46]. These findings were extending on having already identified the utility of ANN in predicting neurological deterioration in mTBI patients (ANN outperformed benchmark CT classification systems, i.e., Marshall, Helsinki, and Rotterdam, in predicting outcomes for paediatric TBI cases) [67]. Notably, a previous study assessing 15 different studies involving 11,026 TBI patients compared various traditional ML models for mortality prediction in moderate and severe TBI patients and found LR to outperform RF, SVM, and ANN [41]. These findings highlight the importance of traditional ML models in clinical data-based studies, with a particular emphasis on ANN, RF, and XGB. Notably, XGB has shown superior performance on clinical data compared to other ML models like DT and ANN. This is likely due to its ability to assess feature importance as effectively as RF, without the need for a preliminary feature selection step [65].

## 3.2 Machine learning using CT images

Table 3 lists the ML studies that have utilized CT images in TBI research applied in both mild and moderate-to-severe patient cohorts. Predominantly the studies used CT images for diagnostic purposes, except Yoon et al. wherein CT scans were used to predict the need for hospital admission, neurosurgical intervention, and 30-day mortality [56]. Twenty out of the 26 studies involved 2D CT images, while the others used 3D volumetric images. This division may arise due to data availability for the various studies; notably 3D CT image volumes are assumed to contain more comprehensive.

A significant trend observed in these studies is the widespread adoption of DL techniques in developing their CT-based ML models. This indicates a clear preference in the field for leveraging the power of advanced AI methods in medical imaging, possibly due to their higher accuracy and efficacy when working with images irrespective of the application area. Interestingly, one recent study did use traditional ML methods [44]. The primary focus of the studies is nonetheless consistent, i.e., concentrating on critical ML tasks including hemorrhage segmentation (or microbleeds), CT image classification, and identifying key findings associated with TBI cases. Regarding TBI severity, the studies present a varied approach: 23 out of 26 focused on moderate-to-severe TBI, likely due to the more apparent and clinically urgent nature of these cases, and two studies encompassed all severities of TBI. Only one study targeted mTBI, a category that often presents unique challenges in diagnosis due to its typically subtle imaging signatures [72].

Nine studies focused on segmenting 2D or 3D CT images. In what can be considered as relatively early work in the field, Scherer used first and second order radiomics statistics, texture, and threshold of image features to segment Intracranial Hemorrhage (ICH) using RF [75]. Concordance coefficient of 0.95 (0.99 for the independent validation set) was achieved between manual segmentation and that produced by the algorithm. Muschelli et al. assessed multiple classifiers for intracranial hemorrhage probability estimation using a rich set of discriminating features, such as CT voxel intensity information. RF chosen to be the best algorithms for this task (Dice similarly with expert at 90%) followed by LR (89%) [47]. Notably, baseline scans from 112 cases could be considered as a small cohort for this type of task. Chilamkurthy et al. analysed over 313,000 head 2D CT scans from approximately 20 centres in India. They demonstrated high capability in detecting a range of intracranial hemorrhages and other critical abnormalities using DL approaches. The findings suggest triaging in the medical setting can be automated [76].. Remedios et al. addressed the challenges of data availability and medical imaging data sharing in their two studies, focusing on the application of multi-site learning. Using data from two different institutions, they demonstrated the superiority of the multi-site model over single-site models. In their first study, they employed an Inception CNN for segmenting hemorrhages and hematomas in 2D CT scans of TBI patients, achieving a Dice Similarity Coefficient (DSC) of 0.64 and a Pearson correlation coefficient of 0.87 [77]. Subsequently, they applied a U-Net architecture for generalized CT hemorrhage segmentation, where the multi-site model achieved an average DSC of 0.69 and a volume correlation of 0.91 [78]. Jain et al. utilized the Collaborative European Neurotrauma Effectiveness Research in TBI (CENTER-TBI) dataset to introduce icobrain, a novel automated method for analysing acute intracranial lesions, cistern volumes, and midline shift on 2D CT images. When benchmarked against expert annotations, this method demonstrated a high intraclass correlation coefficient, showing correspondence rates of 91%, 94%, 93% for acute intercranial lesions cistern volumes, and midline shift, respectively [53]. Building on this foundation, Monteiro et al. utilized the same CENTER-TBI dataset to enhance the application of 3D CNNs for multiclass voxel-wise segmentation of lesion types in TBI patients [74]. Their model demonstrated a high accuracy in quantifying (with an AUC of 89%) and detecting lesions in TBI patients, highlighting its potential in contributing to personalized treatment strategies in TBI. Kuo et al., employed 3D CT scans to introduce Patch fully CNN (PatchFCN) [73]. PatchFCN was trained using 4,396 head CT scans to mimic the analysis performed by neuroimaging radiologists for identifying subtle brain abnormalities. When tested against four radiologists and using 200 head CT scans, the algorithm exhibited remarkable accuracy (AUC = 99%). While Loncaric et al., trained a spatially weighted K-means histogram-based clustering and segmentation approaches along with an iterative morphological processing and volume rendering techniques for ICH segmentation and quantification from digitized CT films. In more than 80% of the cases, the operators concurred with the outcomes of the proposed method [48]. It in fact outperformed two of the four radiologists and demonstrated robust localization of abnormalities, including those missed by experts. These important studies collectively underscore a transformative shift in TBI care, where the integration of advanced DL techniques with head CT scans is not only refining diagnostic accuracy but also paving the way for more nuanced and personalized treatment approaches in TBI management.

**Table 3**: Summary of the studies discussed using CT images with machine learning in TBI.

| Study | Year [Ref] | Model | TBI severity | objective |
|---|---|---|---|---|
| Kuo et al. | 2019 [83] | 3D PatchFCN | Mild to severe | Segmentation |
| Monteiro et al. | 2020 [84] | 3D CNN | | |
| Loncaric et al. | 1995 [61] | 3D TML (WKHC) | Moderate to severe | |
| Scherer et al. | 2016 [85] | 2D TML(FE, RF) | | |
| Muschelli et al. | 2017 [60] | 2D TML ( FE,LR, GAM, RF) | | |
| Chilamkurthy et al. | 2018 [86] | 2D U-Net, NLP, … | | |
| Remedios et al. | 2019 [87] | 2D Inception CNN | | |
| Jain et al. | 2019 [65] | 2D U-Net | | |
| Remedios et al. | 2020 [88] | 2D U-Net | | |
| Ellethy et al. | 2023 [3] | 3D CNN | Mild | Classification |
| Tong et al. | 2011 [89] | 2D TML (FE, SVM, LDA) | Moderate to severe | |
| Keshavamurthy et al. | 2017 [62] | 2D SIFT, CNN, … | | |
| Grewal et al. | 2018 [66] | 3D RADnet | | |
| Jnawali et al. | 2018 [90] | 3D CNN | | |
| Helwan et al. | 2018 [67] | 2D CNN, AE, … | | |
| Ker et al. | 2019 [64] | 3D CNN | | |
| Wang et al. | 2021 [91] | 2D CNN-RNN | | |
| Mushtaq et al. | 2021 [92] | 2D CNN, RNN, … | | |
| Ahmed et al. | 2022 [93] | 2D CNN-LSTM | | |
| Ozaltin et al. | 2022 [63] | 2D OzNet-NCA-ANN | | |
| Pease et al. | 2022 [94] | 2D multi-modal CNN | | |
| Malik & Vidyarthi | 2023 [57] | 2D KNN, SVM, … | | |
| Anjum et al. | 2023 [95] | 2D CNN | | |
| Kadry & Gandomi | 2023 [96] | 2D LDL and SVM | | |
| Yoon et al. | 2023 [68] | 2D CNN, VGG16, … | | |
| Asif et al. | 2023 [69] | 2D Res-Inc-LGBM | | |

*TML: Traditional Machine Learning, FE: Feature Engineering, PatchFCN: Patch based Fully Connected Network, GAM: generalized additive model, NLP: Natural Language Processing, RNN: Recurrent Neural Network, LDL: Light Deep Learning, Res-Inc-LGBM: ResNet101-V2, Inception-V4, and Light Gradient Boosting Machines .*

The recent advancements in ML and DL for brain hemorrhage detection and CT scans classification have been exemplified by a series of studies. Jnawali et al. used 3D CT scans for intracranial hemorrhage classification using CNNs combined with logistic functions and an ensemble approach. This study stands out because of its size, i.e., 40,000 CT scans, and demonstrating an AUC of 87% [80]. Ker et al. expanded on this by applying CNNs for classifying 3D CT brain scans with various hemorrhage types by employing image thresholding techniques to improve classification accuracy [52]. The approach achieved F1 scores between 0.706 and 0.952, proving effective in real hospital scenarios for emergent CT brain diagnoses. We investigated the use of a 3D Multi-Modal Residual CNN with Occlusion Sensitivity Maps for mTBI diagnosis with the TRACK-TBI pilot study dataset [72]. This model showed diagnostic precision improvements with an average accuracy of 82% and an AUC of 95%, when clinical data was combined with 3D CT imaging within our ML framework. Grewal et al. introduced the 3D Recurrent Attention DenseNet (RADnet), a novel approach that aims to emulate the pattern recognition process used by radiologists for analysing CT images [54]. Combining DenseNet architecture with attention mechanisms and a bidirectional LSTM layer, RADnet achieved an 82% prediction accuracy based on 77 brain CTs, surpassing two of the three radiologists in recall.

The detection and classification of brain hemorrhages using 2D CT scans has seen significant advancements with various ML and DL approaches, with early studies looking at shapes and applying SVM to classify shapes and thus differentiate between subdural and extradural hemorrhage (SVM achieving 93% accuracy, versus linear discriminant analysis at 87% accuracy) [79]Kadry & Gandomi focused on a Lightweight DL (LDL) procedure to classify 3D CTs into healthy or haemorrhagic categories [86]. Using a dataset of 2,400 images and employing preprocessing of images using a threshold filter, the LDL was effectively applied for feature extraction, achieving an accuracy rate of over 96% with SVM classification. Wang et al. developed a DL model for acute detection and subtype classification of ICH, and notably winning the 2019-RSNA Brain CT Hemorrhage Challenge with its dataset of over 25,000 scans and high-performance metrics. The model achieved AUCs in the range 98% to 100% across various ICH subtypes [81]. Subsequently, Asif et al. found an accuracy of 98% (sensitivity = 97%; 97% F1 score) using LGBM (combined with ResNet101-V2 and Inception-V4) applied to publicly available CQ500 and RSNA data [57]. Anjum et al. explored brain hemorrhage classification using a lightweight CNN architecture [85]. This study distinguished itself by achieving high accuracy (97%), sensitivity (97%), and specificity (96%) using a relatively small dataset, i.e., 200 images, and applying data augmentation techniques such as rotation, zoom, and horizontal flip. The focus herein was on efficiency and performance, resulting in improved performance over pre-trained models (note, pre-training on different data has been shown to produce good results when application specific data are scarce). Keshavamurthy et al.'s approach combined scale invariant feature transform, 2D CNNs and SVM for rapid and accurate TBI lesion detection. Their method achieved promising results, with a prediction accuracy of 93%, a sensitivity of 91%, and a specificity of 93% [50]. Mushtaq et al. explored the effectiveness of CNN and hybrid models like CNN plus LSTM and CNN plus GRU, for brain Hemorrhage classification. Utilizing a balanced dataset of 200 head 2D CT scans, they achieved 95% detection accuracy [82]. Ahmed et al. also employed a combination of CNN and LSTM to further enhance ICH detection. Their fusion model attained a validation accuracy of 95%, exceeding values reported in prior studies [83]. Ozaltin et al. used OzNet, derived from the classical CNN, and combined with Neighbourhood Component Analysis (NCA) and ANN [51]. The OzNet model was used to create the features from CT images, and the combination of the NCA and ANN algorithms were found to produce the best classification results. Their classification model achieved 99% accuracy in detecting ICH (AUC = 100%, sensitivity = 98%, specificity = 100%; many other combinations achieved high scores as well). In a comprehensively tested study, the application of a CNN for to CT images and clinical data produced a AUC of 82% for an unfavourable outcome as measured by the Glasgow outcome scale, consistent with an AUC of 88% using an internal validation test set [84]. These findings provided a clear suggestion that DL models with head scans can be used to determine six-month outcome after TBI. Malik & Vidyarthi focused on traditional ML models with a large-scale multivariate feature set for classifying brain hemorrhages [44]. Their primary findings were that feature diversity is important for brain hemorrhage detection accuracy. Helwan et al. examined the application of an autoencoder, Stacked Autoencoder (SAE), and 2D CNN models in classifying brain hemorrhages in CT images [55]. They found the SAE model to outperform others, with an accuracy of 91% compared to 88% for the AE and 90% for the CNN model. Yoon et al.'s study effectively integrated algorithmic uncertainty into their DL approach, focusing on the classification of head 2D CT scans and the identification of indeterminate cases. The algorithm they developed was capable of categorizing CT scans based on the probability of ICH or other urgent intracranial abnormalities. Adapted for clinical application following preliminary evaluations, it successfully identified significant intracranial abnormalities, such as hemorrhage, diffuse cerebral edema, and mass effect. This advancement addresses the requirement of identifying neurological abnormalities in clinical settings [56].

## 3.3 Machine Learning Applications in mTBI using other data sources

Table 4 outlines the ML studies in mTBI diagnosis not already identified in the other tables. The categories have primarily been based on the type of data used for the ML prediction. s. Notably, across the different studies a range of ML approaches have been considered, including state-of-the-art classification approaches (i.e., SVM and RF) and ones additionally involving feature generation (e.g., CNN). A broad range of data has been used as well, ranging from clinical data to data collected using a range of medical devices (i.e., MRI, Magnetoencephalography (MEG), and Electroencephalogram (EEG)) and non-medical devices (i.e., eye tracking and speech recordings)

**Table 4**: Summary of machine learning applications using other types of data in mTBI diagnosis.

| Study | Year [Ref] | Model | Data |
|---|---|---|---|
| Vergara et al. | 2017 [87] | SVM | Functional connectivity |
| Bostami et al. | 2022 [88] | SVM, LR, DT, … | |
| Vedaei et al. | 2023 [89] | DSAN | |
| Simos et al. | 2023 [90] | Graph network and ensemble learning | |
| Teng, Mi, et al. | 2023 [91] | RF, SVM, … | |
| Minaee et al. | 2019 [92] | AAE, CNN | Diffusion MRI |
| Ly et al. | 2022 [93] | LR and SVM | Clinical data and MRI |
| Teng et al. | 2023 [94] | SVM | |
| Güler et al. | 2009 [95] | ANN | EEG |
| Lewine et al. | 2019 [96] | ANN, RF, … | |
| Vishwanath et al. | 2021 [97] | CNN, RF, SVM, … | Sleep EEG |
| Aaltonen et al. | 2023 [98] | SVM | MEG |
| Aaltonen | 2022 [49] | LDA, SVM and LR | |
| Tirdad et al. | 2021 [99] | Ensemble learning | Eye movement |
| Wall et al. | 2022 [100] | Bidirectional LSTM | Audio |

*DSAN: Deep Self-Attention Network, LSTM: Long Short-Term Memory, AAE: Adversarial Autoencoder, CRF: Conditional Random Fields, EEG: Electroencephalogram, MEG: magnetoencephalography.*

Numerous studies have considered mTBI diagnosis using an assessment of brain functional connectivity established from rs-fMRI data. Vergara et al. found group independent component analysis was better than SVM in classifying mTBI from healthy controls [87]. The parameters chosen for data preprocessing were found to have a strong influence on classification performance. Bostami et al. suggested that ML implementations which harmonize multi-site mTBI data are required to control for input data variability [88]. In view of this and rs-fMRI producing a range functional connectivity metrics, Vedaei et al. investigated how well different metrics using SVM can classify chronic mTBI [89]. Using Shapley additive explanation analysis of features, they deduced that multi-level seed based functional connectivity measures lead to the best classification performance (AUC = 93%). Simos et al. considered functional connectivity metrics derived from the entire rs-fMRI data (static) and those from a sliding window (dynamic), and reported classification accuracy of 75% (precision = 77%; sensitivity = 74%; specificity = 76%) from the combined use of static and dynamic information [90]. In parallel work, Teng, Mi et al. developed a Deep Self Attention Network (DSAN) framework to classify low and high-order functionally connected networks assessed from rs-fMRI data [91]. They found the combination of low and high-order functional connectivity data was able to achieve the best accuracy (83%) in classifying mTBI, which was a leap in mTBI classification. A subsequent paper by Teng et al. involved a hierarchical feature selection pipeline and considered a range of methods for mTBI diagnosis [94]. They found RF with hierarchical feature selection was able to achieve an accuracy of 90% (precision = 91%, sensitivity = 90%). The small cohort size (i.e., 69 acute mTBI patients and 60 healthy controls) was claimed as a limitation of the study. Minaee et al. found that small datasets can be used effectively when images are converted to patches for mTBI diagnosis from diffusion MRI [92]. rs-fMRI data harmonization across sites, and an appropriate choice of metrics to be classified, are important ingredients within the ML framework used for mTBI diagnosis. The classifier used maybe less important than the input data structure adopted for ML classification. In athletes with concussion, Ly et al. performed a multi-site study involving cognitive measures and MRI scans to classify, using LR and SVM, the presence of concussion [93]. The use of cognitive measures alone produced an accuracy of 74% and a poor sensitivity of 46%. From the MRI data they computed mean diffusivity and fractional anisotropy diffusion measures. With the use of only the mean diffusivity, a similar result was obtained. However, when the cognitive measures were combined with the image-derived metrics, accuracy improved to 74% and sensitivity to 64%. These studies suggest that classification of mTBI can be improved using multi-modal data.

A range of other types of data have also been used to classify mTBI. EEG provides time series data which can be analysed in different ways. Guler et al. used EEG recordings with an ANN to predict the Glasgow outcome scale (in categories of normal, mild, moderate and severe) for the purpose of diagnosing TBI, with a fair agreement between findings of neurologists and ANN prediction [95]. A study by Lewine et al. found global relative theta power (4-8 Hz) increases for mTBI patients, while relative alpha power (8-12 Hz) and global beta-band (12.5-28 Hz) interhemispheric coherence decrease [96]. Overnight EEG recordings classified for mTBI revealed a maximum achievable ML classification accuracy of 95% on the test set, which reduced to 70% when applied to an independent cohort for additional testing [97]. An SVM-based mTBI classification study suggested that theta frequency band identified using MEG is the primary contributor in achieving 79% diagnosis accuracy [98]. MEG power spectra analysis in the range of 1-88 Hz using a range of ML approaches found that mTBI patients can be distinguished from healthy controls with an accuracy in the range 80% to 95% [49]. The authors mentioned that linear ML models may find a place for clinical use, especially to identify patients who are most likely to benefit from close monitoring during the recovery period. The analysis of 3,450 engineered features from saccadic eye movement data led to the identification of 116 features which contribute to mTBI diagnosis with an accuracy of 88% [99]. The complex non-linear pattens in saccadic eye movement were claimed only to be discernible using ML. A different study used the properties of audio recordings of speech collected for mTBI classification [100]. They considered features of the Mel frequency cepstral coefficients and trained a particle swarm optimized bidirectional Long Short-Term Memory (LSTM). The ability to distinguish between mTBI and healthy controls was AUC = 90.4% (sensitivity = 95%; specificity = 86%). Given the vastness of these studies, it is plausible that diagnostic performance of the ML approach can further be improved by combining existing data sources used for classification.

## 4 Discussion

TBI, often referred to as the "silent epidemic," results from external forces disrupting normal brain function. It is a global health concern, with millions of cases reported worldwide. While clinical assessments and CT scans are common and routine practice for diagnosis, their limitations in sensitivity and specificity create challenges, especially in mTBI cases. This review emphasizes the role of ML in addressing these challenges and improving TBI patient care.

This review delineates the distinct roles of ML methods in TBI filed using traditional diagnostic data along with reviewing the ML application in mTBI diagnosis Traditional ML techniques, such as RF and shallow-ANN, are effective in predictive tasks like determining patient outcomes, assessing neurological deterioration, and predicting the necessity of CT scans. These methods working with structured data can provide interpretable models, which are crucial for clinical decision-making. However, their performance is often limited by the quality and dimensionality of the input data.

On the other hand, DL approaches, particularly CNN, show superior performance in image-based tasks. They are adept at analysing complex patterns in imaging data, making them highly effective for detecting hemorrhages, segmenting lesions, and identifying subtle changes in brain imaging that might be indicative of mTBI. While DL methods offer enhanced accuracy and automation in image analysis, they require large datasets for training and may lack the interpretability of traditional ML models.

The integration of diverse data types, including both clinical assessments and neuroimaging data, can lead to more nuanced TBI diagnoses [72], [93]. Employing advanced ML frameworks, such as multimodal DL or ensemble models, can capitalize on the complementary strengths of these varied data sources. This approach enables a more comprehensive analysis, potentially improving diagnostic accuracy and facilitating more personalized treatment strategies for TBI patients.

We explored ML applications based on data sources, categorizing them into clinical data-based models and CT data-based models. Clinical data-based ML models significantly improve TBI patient care by identifying key clinical patterns, aiding in the judicious use of CT scans. This targeted approach not only alleviates healthcare system burdens by reducing unnecessary CT scans and patient radiation exposure but also enables faster and safer discharge of low risk mTBI patients. These advancements highlight the role of ML in enhancing patient profiling, early detection, personalized treatment plans, and resource optimization in healthcare settings, underscoring the transformative impact of ML in medical diagnostics and patient management. A notable limitation of traditional ML methods is that they typically require explicit feature extraction, involving one or more initial steps within the algorithm [101].

CT-based ML models have shown considerable promise in TBI research, with a notable trend towards the adoption of DL techniques. This shift towards advanced AI methods in medical imaging stems from their superior accuracy and efficiency in image analysis tasks. These studies predominantly focus on crucial ML applications such as hemorrhage segmentation, CT scan classification, and detecting key indicators in moderate-to-severe TBI cases. CT scans are particularly effective in identifying the more pronounced structural damage associated with moderate to severe TBI, offering clear and interpretable imaging of such injuries [101], [102]. Notably, CT scans were not utilized for developing TBI prognostication ML models, and they were employed in only one study to deploy a diagnostic model for mTBI. This distribution of research focus underscores the need for more attention and research efforts in the domain of mTBI, given its prevalence and clinical significance. Expanding the development of CT-based ML models for mTBI diagnosis and prognosis could be highly beneficial in improving patient care in this specific TBI category.

Figure 3 illustrates a taxonomy of ML models used in TBI research. The tree structure organizes the ML models into branches according to the data types they are typically associated with. The 'CT Scans' branch at the top-right is dedicated to image-based models, prominently featuring algorithms like Inception CNN and U-Net. Opposite this, the 'Clinical Data' branch at the top-left lists models such as ANN and RF, which are widely employed with clinical datasets. The 'Other Data Source' branch captures the breadth of models adaptable to diverse datasets, including AAE and LSTM. Notably, certain models like ANN, RF, SVM, and LR are versatile enough to be applied regardless of data type. CNN, LSTM, and LDA models find common ground between CT-based data and other integrated data sources such as rsfMRI. Similarly, DT and ensemble learning bridge the gap between models used for clinical data and those utilizing other data sources like EEG.

MRI-based ML models are revolutionizing TBI diagnostics by utilizing MRI's superior soft tissue imaging capabilities. These models are particularly adept at detecting intricate details like microhemorrhages, diffuse axonal injuries, and small areas of scarring or contusions, which are often undetectable by CT scans [103], [104]. Studies have shown that a significant portion of mTBI patients with normal CT scans exhibit TBI-related abnormalities in MRI scans [19]. The integration of MRI data with advanced ML techniques, including fMRI and diffusion MRI, has resulted in more accurate and sensitive models in mTBI diagnosis. While MRI is not the standard clinical neuroimaging test for TBI, its combination with ML techniques significantly enhances our understanding and management of TBI, offering new insights into this complex condition.

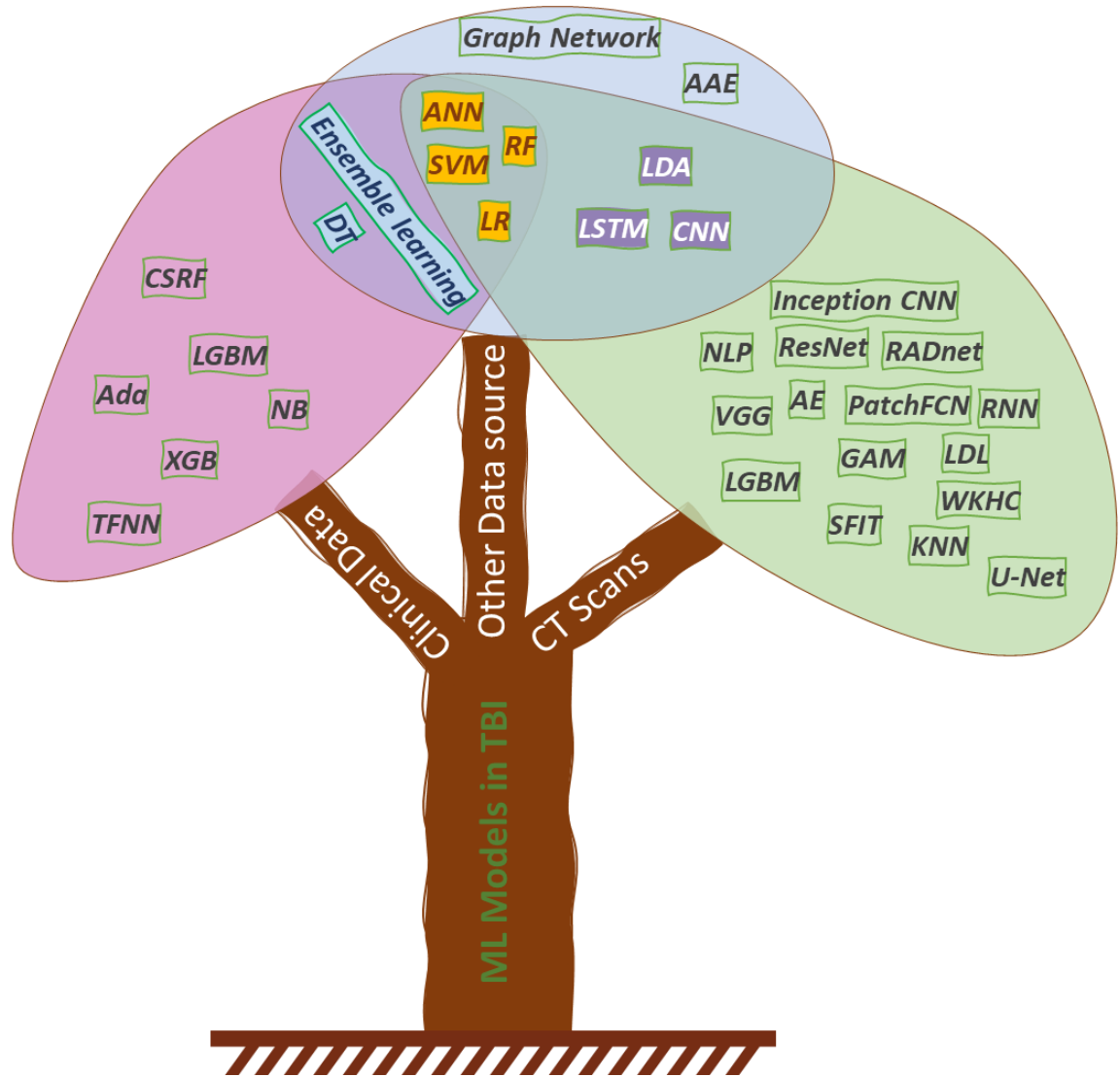

**Figure 3: An overview of ML models utilized in TBI research, categorized by data source.**

ML: Machine Learning, DL: Deep Learning, ANN: Artificial Neural Network, LR: Logistic Regression, NB: Naive Bayes, DT: Decision Tree, KNN: K-Nearest Neighbour, SVM: Support Vector Machine, XGB: Extreme Gradient Boosting, RF: Random Forest, WKHC: Weighted K-means Histogram-based Clustering, LDA: Linear Discriminant Analysis, SIFT: Scale Invariant Feature Transformation, NCA: Neighbourhood Component Analysis, CNN: Convolutional Neural Network, RADnet: Recurrent Attention Dense Network, AE: Autoencoder, VGG: Visual Geometry Group, ResNet: Residual Neural Network.

As shown from the analysed studies, seven studies employed traditional diagnostic data for mTBI. Five studies used clinical data to enhance the decision-making process in diagnosis, one study utilized clinical data along with the CT reports to confirm mTBI cases, and one another used CT scans integrated with clinical data for analysis and diagnosis of mTBI. However, fifteen studies used advanced MRI modalities and other data sources for mTBI diagnosis. Despite current advancements, mTBI diagnosis still presents challenges ripe for further exploration. A key aspect is the integration of standard neuroimaging data, either alone or in conjunction with multimodal data. This involves merging patient clinical data with imaging data to create more comprehensive and accurate diagnostic models. Additionally, there is a growing need for ML models capable of adapting to the variability in imaging techniques and image quality arising across healthcare settings. Equally important is the improvement of the interpretability of DL models, ensuring they are robust and generalizable across different patient populations. Addressing these areas is crucial for enhancing the efficacy of mTBI diagnosis.

While the advancements in ML for TBI diagnosis are promising, they come with their own set of challenges. Key among these is the difficulty in gathering sufficient data from single sites, necessitating algorithms capable of handling indeterminate imaging findings and significant uncertainty. Additionally, ethical considerations, including patient privacy and consent, are paramount in the application of ML in healthcare and require thorough attention.

## 5 Conclusion

In summary, traumatic brain injury (TBI) represents a significant global health challenge with profound impacts. This review underscores the crucial role of machine learning (ML) in advancing the diagnosis and prognosis of TBI, particularly in mild TBI. ML models, encompassing both traditional algorithms and deep learning (DL) techniques, offer innovative approaches to address TBI-related challenges. These models enhance patient care by leveraging clinical and neuroimaging data, aiding in early detection, and potentially facilitating changes in personalized treatment strategies. Traditional ML methods prove effective in tasks such as predicting patient outcomes and the necessity for CT scans. While DL techniques demonstrate their strength in image analysis, excelling at detecting hemorrhages and segmenting lesions. The integration of ML with diverse data sources, including clinical assessments and neuroimaging, promises significant advancements in TBI field. The studies reviewed show high accuracy and efficacy in image analysis, indicating a preference for advanced AI methods in medical imaging. As these models continue to develop and be refined for specific applications, the future of TBI diagnosis and treatment looks increasingly accurate, efficient, and personalized. These developments highlight the potential of ML in providing comprehensive patient profiling, early detection, personalized treatment plans, accurate outcome prediction, efficient resource allocation, and advancing research and clinical decision-making support.